# Evolution of value-based decision-making preferences in the population

Erdem Pulcu[1]*

**Affiliations:**

[1]University of Oxford, Department of Psychiatry, Computational Psychiatry Lab, OX3 7JX UK

*Correspondence to: Dr. Erdem Pulcu, erdem.pulcu@psych.ox.ac.uk

**Abstract:**

**We are living in an uncertain and dynamically changing world, where optimal decision-making under uncertainty is directly linked to the survival of species. However, evolutionary selection pressures that shape value-based decision-making under uncertainty have thus far received limited attention. Here, we demonstrate that fitness associated with different value-based decision-making preferences is influenced by the value properties of the environment, as well as the characteristics and the density of competitors in the population. We show that risk-seeking tendencies will eventually dominate the population, when there are a relatively large number of discrete strategies competing in volatile value environments. These results may have important implications for behavioural ecology: (i) to inform the prediction that species which naturally exhibit risk-averse characteristics and live alongside risk-seeking competitors may be selected against; (ii) to potentially improve our understanding of day-traders' value-based decision-making preferences in volatile financial markets in terms of an environmental adaptation.**

**Introduction:**

We are living in an uncertain and an ever-changing world, where our decisions are guided by our expectations of their outcomes. Optimal decision-making under uncertainty is a common problem faced by all biological entities in higher classes of the animal taxonomy, and it is crucial for the survival of species. The ways in which we perceive probabilities associated with desirable or aversive outcomes is an important factor in shaping our expectations, and a key tenet of modelling [reinforcement] learning and value-based decision-making processes (Behrens, Woolrich et al. 2007). Consequently, decision-making under uncertainty has been studied extensively in economics (Kahneman and Tversky 1979, Tversky and Fox 1995, Prelec 1998), as well as in behavioural and neural sciences (Hsu, Bhatt et al. 2005, Tobler, O'Doherty et al. 2007, Hsu, Krajbich et al. 2009, Hunt, Kolling et al. 2012); aiming to shed light on how the brain extracts relevant information from the environment to resolve uncertainty, in order to make decisions optimally.

Although theories of value-based decision-making are continuously expanding to account for various non-normative aspects of human behaviour observed in field experiments, historically two theories have been particularly influential: *Expected Utility Theory* (Mongin 1997, Dhami 2016) and *Prospect Theory* (Kahneman and Tversky 2013). Prospect Theory is often regarded as an advancement over the Expected Utility Theory as it accounts for perception of risk as well as non-linear modulation of outcome probabilities, an aspect of value-based decision-making which is often regarded as suboptimal (Allais 1990). Despite their significance, the impact of evolutionary biological pressures shaping behavioural traits such as risk perception and non-linear probability weighting which govern value-based decision-making, have thus far received almost no empirical attention (Sinn 2003, Santos and Rosati 2015). The empirical studies on these aspects are particularly scarce in comparison to game theoretic (Smith 1993, Von Neumann and Morgenstern 2007, Camerer 2010) and social interactive processes such as interpersonal cooperation (Nowak and Sigmund 1993, Axelrod 1997) or altruistic punishment (Boyd, Gintis et al. 2003).

The current manuscript addresses this knowledge gap by bridging the stochastic choice and stochastic population models in an evolutionary/computational biological framework: providing quantitative analyses of fitness trajectories associated with different value-based decision-making strategies competing against each other in dynamically changing environments. The macroscopic approach presented here is important not only because the global financial markets, where millions of traders interact every day, remain just as volatile as the physical environment of the Prehistoric times; but also, to understand the role of evolutionary biological pressures which shape factors (e.g. risk attitude and probability weighting) that influence value-based decision-making preferences in the population.

# Results

## *Optimal strategy in the deterministic choice model*

In order to study how evolution might have shaped attitudes to value-based decision-making in the population, we conducted a series of simulated decision-making experiments (Fig. 1A), in which agents were defined in terms of their risk attitudes (i.e. risk averse, risk neutral and risk seeking) and probability weighting preferences (i.e. unbiased, probability overweighting, probability underweighting, S-shaped and inverse S-shaped; see Supplemental Materials and Methods (SMM) for mathematical definitions; and Supplementary Figure 1 for their graphical expression) and all of their 15 possible categorical combinations; competing in a virtual environment containing 1 million randomly generated options, from which rewards are delivered probabilistically (Fig. 1). The mathematical models in the current paper build up on Stott 2006, who, after fitting 256 combinations of different risk, probability weighting and stochastic choice functions, recommended the use of a power utility function, a probability weighting function based on the work of Prelec (1998), and a Logit function (Eq.1) for value-based decision-making.

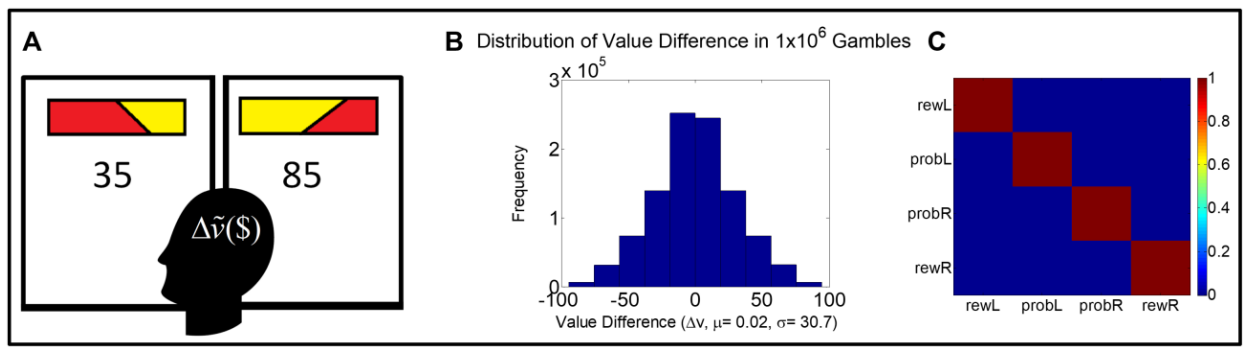

**Fig. 1**. **Decision-making in probabilistic gambles and the value properties of the simulation environment.** **(A)** Schematic diagram of the value-based decision-making experiment. **(B)** Histogram for the distribution of the expected value difference between the options ($\Delta v$) across 1x10⁶ probabilistic gambles. **(C)** The reward magnitudes and probabilities for each option were decorrelated. The colour bar shows the correlation coefficient (*r*).

As one might expect, when the agents make decisions in isolation, those employing the risk neutral strategy with unbiased probability weighting accumulated more resources relative to the other strategies when the expected value difference between the options ($\Delta v$) in the environment varied randomly in a wide range (Supplementary Fig. 2A), or within a limited range ($|\Delta v| < 5$; Supplementary Fig. 2B). These initial simulations in which the agents, hypersensitive even to the subtlest changes in the expected value difference between the options, make decisions in isolation set the benchmark in favour of the risk neutral strategy with unbiased probability weighting under the deterministic choice model. However, behavioural economic experiments show some degree of stochasticity in people's value-based decisions (Hsu, Krajbich et al. 2009).

*Optimal strategy in the stochastic choice model*

In mathematical models of decision-making, the degree of stochasticity is defined by an inverse temperature term ($\beta$), adopted from thermodynamics (also see SMM). Assigning a moderate value to the $\beta$ coefficient, which modulates the subjective value difference between the options ($\Delta \tilde{v}$) in a Logit function (which in return generates the choice probabilities of each of the available options (Daw 2011)):

$$q_L = 1/(1+\exp^{(-\beta(\Delta \tilde{v}))}) \qquad (1)$$

suggests that choice stochasticity will have a negative effect on the performance of the different strategies, particularly on those with an element of probability underweighting (Supplementary Figure 2C). Furthermore, by using the risk neutral strategy with unbiased probability weighting as reference, we demonstrate that increasing values of the $\beta$ coefficient quickly saturates the magnitude of accumulated rewards (Supplementary Figure 2D), potentially indicating the upper boundary of its evolution in the population if it is also subjected to selection as a behavioural trait.

A formal statistical analysis conducted on the amount of rewards accumulated by these 15 categorical strategies suggested a significant main effect of risk attitude ($F(2,19998)=19529$, $p<.001$), a significant main effect of probability weighting ($F(4,39996)=33458$, $p<.001$) and a significant interaction term ($F(8,79992)=27736$, $p<.001$; Supplementary Figure 2E).

***Optimal strategy in the stochastic population model***

Following this rather necessary introduction, we progress with a population level of analysis by duplicating the agents from the first stage to build up a mixed, model society ($N=4.5 \times 10^4$) in which each of these 15 different strategies occupied an equal population density.

We created volatile value environments by segmenting the original 1 million gambles into 10,000 evolutionary time courses each running for 100 generations, where the expected value difference between the options changed randomly from one generation to the next. The reward magnitudes in each probabilistic gamble corresponded to the amount of resources which can be acquired from the physical environment during the course of a single generation on the simulation timeline (Fig. 2A). Although this abstraction reduces dimensionality of the complexity of our everyday value-based decisions, it would still capture the influence of such important and sequential decisions (e.g. deciding to purchase a property which is 80% likely to appreciate in value) on one's reproductive fitness. For example, hallmark evolutionary biological studies of interpersonal cooperation (Nowak and Sigmund 1993, Axelrod 1997) also rely on experimental paradigms abstracting such social behaviours over a predetermined matrix game (e.g. the Prisoner's Dilemma Game), where agents interact iteratively. Creating 10,000 randomly generated environments helps us to capture the evolutionary/macroscopic picture over many simulation environments with different value properties: those which change gradually, as well as those which are highly volatile.

We linked the individual stochastic-choice model with an evolutionary dynamic computational model (i.e. a stochastic population model) at the expected value ($\Delta v$) and the choice probability ($q_L$) levels (see SMM for the full mathematical description of the models); making it possible to compute the expected random fitness ($F_A$) for any of the 15 aforementioned strategies competing against each other to acquire rewards, and reproduce. Once the expected random fitness ($F_A$) of each of the competing strategies is computed, it is possible to model the local process of co-evolution by natural selection, as previously proposed by Traulsen et al (Traulsen, Claussen et al. 2005). Natural selection is implemented in terms of bidirectional stochastic transition rates between the groups ($r_{A \to B}$) from one generation to the next and these are proportional to between-group differences in expected random fitness.

In contrast to the results of the individual choice models where agents make decisions in isolation, the stochastic population model reveals that increasing the value of the $\beta$ coefficient enhances the performance of the risk neutral strategy with unbiased probability weighting, whereby it acquires higher population density at time ($t$) = 100 (the time point where each simulation ended (Fig. 2B~G)), as well as improving the overall fitness of the population (Supplementary Fig. 3B). Here, it is important to point out that across all the simulations using the stochastic population model reported in this manuscript, the upper boundary of the $\beta$ coefficient was set to 2.6 which was previously reported by Hsu et al.(Hsu, Krajbich et al. 2009) in the context of value-based decision-making.

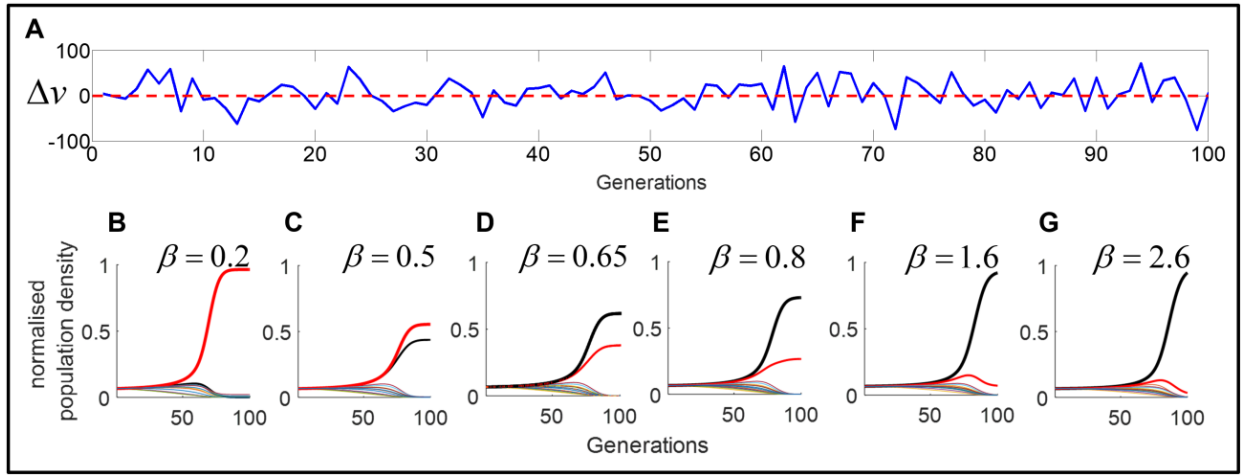

Fig. 2. Evolution of value-based decision-making preferences in volatile environments. (A) Graphical expression of a single simulation environment with respect to expected value difference between the options ($\Delta v$), where each gamble is treated as a generation on the evolutionary timeline. (B-G) Increasing values of the β coefficient improves the evolutionary fitness of the risk neutral strategy with unbiased probability weighting (thick black lines in panels D~G), while the evolutionary fitness of all the other competitors are negatively affected. Remarkably, lower values of the β coefficient favour the risk-seeking strategy with unbiased probability weighting (thick red lines in panels B and C).

### *Is risk neutral strategy with unbiased probability weighting an Evolutionarily Stable Strategy (ESS) for value-based decision-making?*

The competition in the population is shown to be the strongest when $\beta \simeq 0.55$ and the risk neutral strategy with unbiased probability weighting appeared to be the most optimal strategy overall (see Supplementary Figure 3). From an evolutionary fitness point of view, it is noteworthy that a previous behavioural study also reported parameter values for a 2-parameter probability weighting function which would correspond to unbiased probability weighting in the gains domain (Stott 2006). However, unlike behavioural studies which focus on participants' choices in isolation, our population level analyses show that the performance of the risk neutral strategy with unbiased probability weighting also depends on how choice stochasticity influences the performance of other competing strategies, therefore it cannot be an Evolutionary Stable Strategy on its own (ESS; (Smith 1982)).

***The effect of environmental volatility on the fitness of different value-based decision-making preferences***

When $\beta \simeq 0.55$ and the competition between 15 categorical strategies is strongest, only 4 strategies eventually dominated the population (% out of 10000 simulations in brackets): risk neutral (49.79%) and risk-seeking strategies with unbiased probability weighting (48.23%), risk-seeking strategy with probability underweighting (0.0032%) and risk-seeking strategy with an S-shaped probability weighting function (0.0016%). It is important to note that across 10,000 different simulation environments, all other strategies are consistently driven to extinction (Supplementary Figure 4). In order to understand how the nature of environmental volatility influence the fitness of these successful strategies, we considered three quantitative measures: magnitude of change in the expected value of the environment from one generation to the next; the frequency of the change in the sign of expected value difference (from positive to negative, or vice versa); and how gradually the expected value difference changed in the environment by checking the correlation coefficient between a vector containing the number of generations and a vector containing expected value differences. Here, a highly positive or a highly negative value for the correlation coefficient would mean that environment, although volatile, is changing relatively more gradually from one generation to the next (e.g. similar to bear or bull markets; (Gonzalez, Powell et al. 2005)). Subsequent analysis suggested that risk-seeking strategy with S-shaped probability weighting function prevailed in environments in which the average magnitude of the change in the expected value from one generation to the next was highest (see Fig 3; $F_{3, 9846}$=3.473, p=0.015). The environments where different strategies eventually dominated the population, were comparable with respect to other metrics of volatility (all $F_{3, 9846}$ < 1.9716; all p> 0.115).

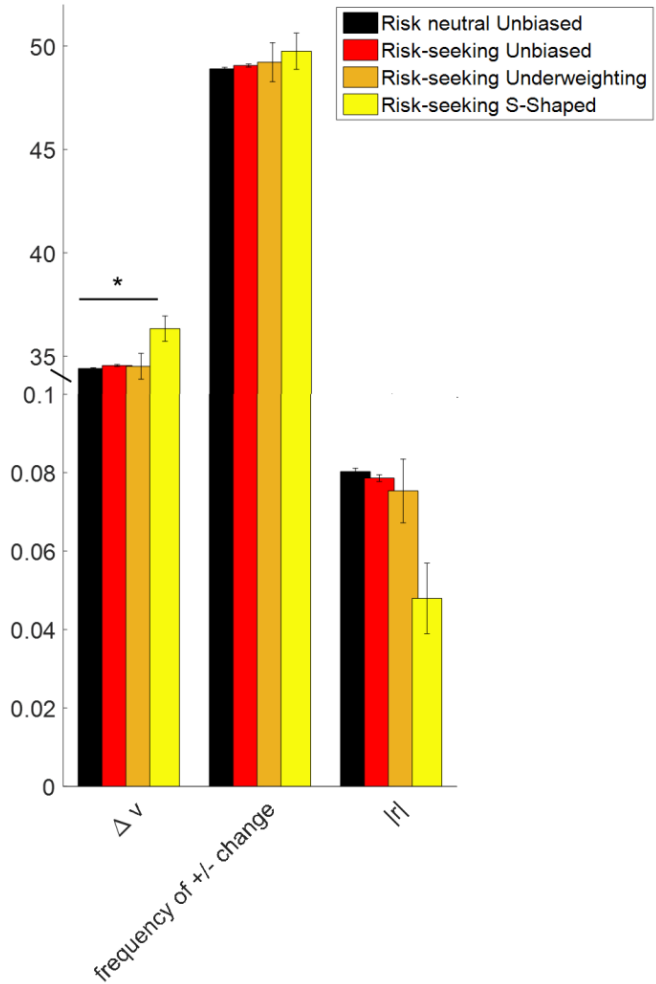

**Figure 3. The relationship between different metrics of environmental volatility and successful strategies.** A formal statistical analysis suggests that environments in which the change in the expected value difference from one generation to the next is highest favours risk-seeking strategies with an S-shaped probability weighting function (*p<0.05).

***Co-evolution of value-based decision-making preferences in populations with inherent variability***

It is assumed that behavioural traits with large variability observed in the population, such as value-based decision-making (Hunt 2014), emerge, carry on existing concurrently, and have co-evolutionary dynamics. However, it is not possible to argue that 15 categorical strategies that we initially defined can capture all the individual variability in value-based decision-making preferences that one can observe in a behavioural field experiment. To address this omission, we wanted to investigate the evolution of value-based decision-making preferences in a population with a large degree of inherent variability where transient competitors, those which fall outside of our predetermined/categorical strategies, occupied equal population density.

In order to generate these transient strategies, we varied the values of the α and the ρ coefficients in two probability weighting functions and the power utility function simultaneously (see SMM; Eq. 8 and 1, respectively) on a 12x40 numerical grid (i.e. the parameter space). Here, varying the values of the α coefficient produced strategies with different probability weighting preferences (see SMM for mathematical definitions and Fig. 4A for a graphical expression of the probability weighting functions); whereas varying the values of the ρ coefficient from 0.5 to 1.5 covered possible degrees of risk perception, in total producing 480 discrete strategies. Then, the simulations described above were repeated in the same 10,000 volatile environments and the average normalised population density of the unbiased strategy at each intersection was then converted to a heat map (Fig. 4B).

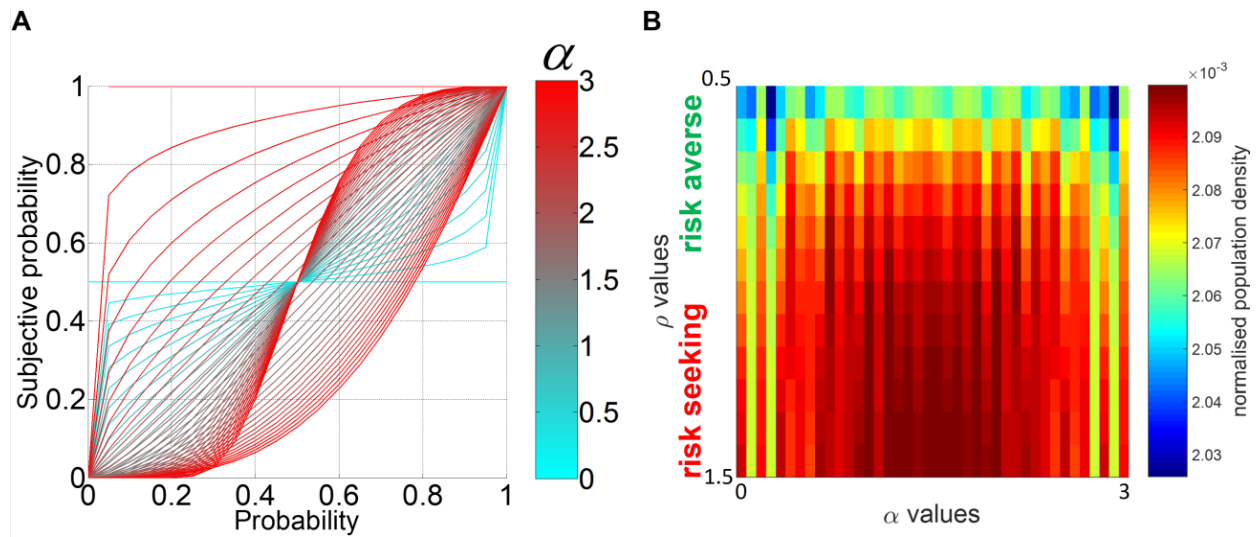

**Figure 4. Co-evolution of value-based decision-making in populations with inherent variability. (A)** The graphical expression of the way the objective probabilities were modulated at each step of risk perception, produced by varying the values of the α coefficient in Equations 8, simultaneously. The colour bar shows the values of the α coefficient in the probability weighting functions. **(B)** The heat map shows the average normalised population density of 480 competing strategies at $t = 100$, averaged across 10,000 simulations. Values of the risk parameter ($\rho<1$) indicate risk aversion and ($\rho>1$) risk-seeking preferences, respectively.

This investigation showed that the performance of the risk neutral strategy with unbiased probability weighting is sensitive to the changes in the characteristics of the competitors in populations with inherent variability. The heatmap of normalised population densities (Fig 4B), as well as a complementary simulation which was run for 3000 generations until the population reaches an equilibrium (Supplementary Video), clearly highlights an area where natural selection favours risk-seeking strategies relative to their competitors (parameter combination for the best strategy: $\rho= 1.5$; $\alpha= 1.3333$ in Eq. 8a, also see Supplementary Figure 5 in which we show that simulations in 10,000 different environments discriminate between the best and the second-best strategies).

**Discussion**

The present results demonstrate that evolutionary fitness associated with different value-based decision-making strategies is influenced by agents' choice stochasticity (Fig 2B-G), seemingly subtle differences in the value properties of the environment (Fig 3), and the characteristics and the density of competitors in the population (Fig 4B). Although having a risk neutral attitude while maintaining an unbiased perception of outcome probabilities is commonly regarded as the optimal policy, the numerical analysis provided here suggests that agents' risk attitude interact with their probability weighting preferences in shaping their overall fitness. As a result, the global evidence suggests that the risk neutral strategy with unbiased probability weighting cannot be an ESS on its own. The agent-based stochastic population models show that natural selection favours risk-seeking preferences when there are relatively high number of discrete strategies competing in the population (Fig. 4B and Supplementary Figure and Supplementary Video).

*Potential implications for behavioural ecology*

The macroscopic/evolutionary approach which is presented here may provide valuable insights for behavioural ecology. Formulating population models of value-based decision-making preferences over risk perception and probability weighting dimensions, is critical for developing a canonical understanding of decision-making processes in predator-prey encounters (Lima 2002, Hebblewhite, Merrill et al. 2005), during foraging considerations (Orrock, Danielson et al. 2004, Higginson, Fawcett et al. 2012) and the trade-offs between them (Hebblewhite and Merrill 2009), all of which are related to evolutionary fitness and natural selection of species. Laboratory studies of probability weighting and risk decision-making can inform the development and fine-tuning of these population models.

For example, it is known that higher order primates are capable of tracking probabilities associated with pleasant as well as undesirable outcomes (Lakshminarayanan, Chen et al. 2011); with probabilities associated with rewards being encoded in the midbrain dopaminergic (Fiorillo, Tobler et al. 2003) and posterior cingulate neurons (McCoy and Platt 2005). A recent study showed that computations underlying probability weighting in monkeys also utilise nonlinear functions (Stauffer, Lak et al. 2015), similar to those which account for human behaviour. Utilising the 1-parameter probability weighting function, which is in essence a modified version of the 2-parameter function (Prelec 1998) obtained by setting the value of the δ parameter to 1, the authors showed quantitative similarities between humans (ϒ = 0.74; (Wu and Gonzalez 1998, Tanaka, Camerer et al. 2010)) and monkeys in terms of probability weighting (ϒ = 0.31 and 0.47 in two different experimental animals, respectively). These empirical studies highlight that monkeys overweigh probabilities below 0.35 to a relatively greater extent than humans. However, the best strategy for value-based decision-making (which we identified here) is not only a function of probability weighting, but also risk perception. The best strategy is shown to underweight probabilities across the probability spectrum (Supplementary Figure 6).

On the other hand, there is accumulating evidence in favour of variability in risk decision-making across species. For example, bonobos (Heilbronner, Rosati et al. 2008) and lemurs (MacLean, Mandalaywala et al. 2012) show risk aversion, whereas rodents (Adriani and Laviola 2006) and macaques have preference for risky options (McCoy and Platt 2005, Hayden and Platt 2007). Although it is not known how well these laboratory findings could represent computations underlying value-based decision-making in the wild (Paglieri, Addessi et al. 2014), one clear prediction of our model is that when there are large number of species competing to acquire resources in finite and volatile environments, those with pronounced risk aversion could eventually be selected against.

***Potential implications for understanding risk aversion in the population***

Inevitably, this prediction raises questions about the prevalence of risk aversion in the population, which is consistently observed in human participants (Cohn, Lewellen et al. 1975, Kahneman and Tversky 1979, Pulcu and Haruno 2017). Our quantitative analysis suggests that the fitness of value-based decision-making strategies depend on an interaction between risk perception and probability weighting (Fig 4B); as well as, at the macroscopic level, the value properties of the environment (Fig 3). This means that the evolutionary fitness of strategies that include risk aversion is also influenced by probability weighting preferences (e.g. the variability in the upper half of Fig 4B). Therefore, an evolutionary/future-guided prediction in favour of risk-seeking preferences is not necessarily in conflict with the results of existing behavioural studies, as evolutionary computational models operate on the infinite timeline. Although studies revealing the relationship between key value-based and social decision-making traits in computational terms are lacking, it is possible that risk aversion survives in the population through these means, particularly if it is related to behavioural traits which might give it a fitness advantage over other competing strategies in the population (e.g. interpersonal cooperation).

***Potential implications for understanding non-linear probability weighting preferences***

As we highlighted previously, a considerable number of studies in which human participants choose between probabilistic rewards under uncertainty have reported probability weighting preferences with nonlinear properties: an overweighting tendency for probabilities approximately lower than 0.35, but a marked underweighting for probabilities exceeding this threshold (Stott 2006, Hsu, Krajbich et al. 2009, Tanaka, Camerer et al. 2010) (also see Supplementary Figure 6). On the other hand, the studies which used a probability weighting function similar to the $\log_2$ functional form reported here focused on how people make value-based decision while learning the hidden probabilities associated with rewards or punishments by predictive sampling in volatile environments (Behrens, Woolrich et al. 2007, Suzuki, Harasawa et al. 2012, Browning, Behrens et al. 2015). Arguably, these latter experimental designs may have higher ecological validity in terms of understanding probability weighting preferences in the population in real life financial decision-making situations, considering that decision-makers do not always have full access to decision variables necessary for computing the expected value difference between the options they face. The probability weighting functions reported by the latter studies also have nonlinear properties to account for their subjective modulation, but unlike the previous studies mentioned at the beginning, their functional form was mainly expressed in terms of an underweighting tendency for probabilities lower than 0.5 and an overweighting tendency beyond this cut-off point (Behrens, Woolrich et al. 2007, Suzuki, Harasawa et al. 2012). In the current work, we provide evidence showing that under favourable conditions (Fig. 3), risk-seeking individuals who utilise a similar probability weighting function to guide their value-based decisions in volatile environments/markets will be the most competitive agents in terms of evolutionary fitness, particularly in environments where the volatility is high in terms of the magnitude of the change in the expected value of resources from one generation to the next. Taken together, our results suggest that the novel $\log_2$ functional form reported here, may be another suitable candidate to represent probability weighting preferences in humans for everyday financial decision-making.

***Potential implications for understanding risky decisions in financial markets***

From a complementary perspective, the findings we present here may have important implications for understanding day-traders' decisions in global financial markets, also including those which involve cryptocurrency exchanges, which exhibit similar volatile characteristic to those of our simulated environments. It is frequently debated whether risky decisions are among the triggering causes of global financial crises (Rajan 2005), which seem to have shortening cycles.

Here, we showed evidence to suggest that volatile financial markets in which traders are expected to make such value-based decisions rapidly and sequentially, where the stakes are high and poor performers are eventually eliminated, may produce more favourable outcomes for those with risk-seeking tendencies. Therefore, it is possible to think of risk-seeking tendencies observed in these populations in terms of an evolutionary/environmental adaptation. Our results highlight an overarching evolutionary biological mechanism, complementing the findings of previous studies which showed neural computations underlying how observing others' value-based decisions could influence one's own preferences in the same direction (Chung, Christopoulos et al. 2015, Suzuki, Jensen et al. 2016). The present results are also in line with the predictions of a seminal work which raised the possibility that chronic exposure to cortisol in response to the volatility of financial markets could shift one's risk preferences (Coates and Herbert 2008). Taken together, these endocrinological, neural and population level mechanisms may lead to spread of risk-seeking tendencies in day-traders exchanging in competitive financial markets.

***Potential implications for understanding evolutionary biological roots of vulnerability to behavioural pathologies***

Finally, our macroscopic approach could also inform the evolutionary perspective on psychopathology (Baron-Cohen 2013), which posits that clinically debilitating conditions may survive in the genetic selection pool because of their associated fitness advantages. We propose that risk-seeking strategies with different degrees of probability weighting, which could be highly adaptive when agents are competing for finite resources in volatile environments, may contribute to a hardwired, biological vulnerability feature for psychiatric disorders associated with risk and sensation seeking behaviours; such as pathological gambling (Clark 2010).

*Acknowledgments:*

We would like to thank Dr. Shinsuke Suzuki, Dr. Lalin Anik, Dr. Michael Browning, Prof. Masahiko Haruno and Ms. Alexandra Pike for their helpful comments on an earlier version of this paper.

**Supplementary Materials:**

Supplementary Methods

Figures S1-S6

Video

References

**Supplementary Methods:**

### 1. The probabilistic gambles

We have generated $1 \times 10^6$ probabilistic gambles by using MATLAB's *randsample* function. The reward magnitudes ranged between 10 and 100 with 5 point increments, and the probabilities ranged between 0.05 and 0.95 with 0.05 increments. The expected value difference ($\Delta v$) between the left ($\pi_L$) and the right ($\pi_R$) gambles had a mean value 0.02 and standard deviation 30.7 (see Figure 1B; also see below for the notations). Reward magnitudes and probabilities were generated to be decorrelated (see Figure 1C).

### 2. Definition of risk attitudes

In the current study, the risk attitudes in value-based decision-making were captured by a power utility parameter ($\rho > 0$), where $\rho < 1$ indicate risk aversion, $\rho = 1$ indicate risk neutrality and $\rho > 1$ indicate risk seeking preferences. The expected utility of the reward magnitudes (*m*) were computed as follows

$$U = m^\rho \tag{1}$$

### 3. Definition of different probability weighting preferences

The strategies which underweighting and overweighting probabilities were defined by the 2-parameter probability weighting function(Prelec 1998):

$$\tilde{p} = \exp^{(-\gamma(-\ln(p))^\delta)} \tag{2a}$$

where parameters $\gamma$ and $\delta$ were set to 3 and 1.05 for the underweighting (UW); and 0.5 and 1.05 for the overweighting strategy (OW), respectively.

We also considered two hybrid strategies which had shifting probability weighting preferences. For example, the S-shaped (S-S) strategy under weighs probabilities less than 0.5, and overweighs probabilities more than 0.5. It acts comparably with the unbiased (UB) strategy when the probability is 0.5 and its probability weighting function is defined by the formula:

$$\tilde{p} = 2^{(-(-\log_2(p)^\alpha))} \qquad (2b)$$

where the parameter α is set to 3. We also considered inverse S-shaped probability weighting preferences, for which the parameter α is set to 0.5.

Irrespective of the value of the α parameter, the $\log_2$ functional form always crosses the p/p diagonal at 0.5 and consequently accurately captures the intuition that, psychologically, wide majority of people will have an unbiased perception of the 50/50 odds.

The graphical expression of different probability weighting preferences is summarised in Supplementary Figure 1.

**4. Individual stochastic-choice model**

One can think of 15 unique combinations of categorical risk attitudes (e.g. risk seeking) and probability weighting preferences defining different approaches to value-based decision-making. Agents adopting any of these strategies compute the expected value of a gamble they face accordingly:

$$\pi = U\tilde{p} \qquad (3)$$

and make their choices in relation to the subjective value difference between each gamble (i.e. here, the difference between left and right options):

$$\Delta\tilde{v} = \pi_L - \pi_R \qquad (4)$$

trial-wise stochastic choice probabilities of each option follow Luce's choice axiom and choice probabilities for each gamble are generated by a sigmoid function (Daw 2011):

$$q_L = 1/(1+\exp^{(-\beta(\Delta \tilde{v}))}) \quad (5)$$

where $\beta > 0$, is the inverse temperature term adopted from thermodynamics and it determines the degree of stochasticity in choice probabilities; values of $\beta \to 0$ giving way for stochastic choices, and values of $\beta \to \infty$ leading to deterministic choices.

## 5. The stochastic population model

After defining the stochastic choice model for the value-based decision-making at the individual level, we constructed a stochastic population model by applying a kinetic Monte Carlo algorithm(Gillespie 1976), to define the local process of the evolutionary game(Traulsen, Claussen et al. 2005, Bladon, Galla et al. 2010). At time $s \in [0, T_{end}]$, we defined the random populations of each of the subgroups as $N_A(s)$.

The local process of the between-group competition (Traulsen, Claussen et al. 2005) is then defined accordingly:

$$N_A \to N_A - 1, N_B \to N_B + 1, \text{ with a rate of } N \cdot r_{A \to B} \quad (6a)$$

applicable for all possible combinations of $A$ and $B$ (i.e. covering all possible transitions between the subgroups).

Here, $N$ is the constant population size of the system (i.e. a linear Moran process(Bladon, Galla et al. 2010)), which is fixed to $4.5 \times 10^4$ agents with each of the competing groups occupying $1/15^{th}$ of the population for an unbiased investigation of their evolutionary fitness.

The expected random fitness of any agent in group $A$ is defined by

$$f_A := q_L^A \cdot \pi_L + (1 - q_L^A) \cdot \pi_R \qquad (6b)$$

The transition rates between the groups, $r_{A \to B}$'s, are then defined by the formula:

$$r_{A \to B} := \frac{1}{2} \frac{N_A}{N} \frac{N_B}{N} (1 + \frac{F_B - F_A}{\Delta F_{max}}) \qquad (6c)$$

In this formulation, $F_A$ is the random (expected) fitness of the group $A$:

$$F_A := f_A \cdot N_A \qquad (6d)$$

It is important to point out once again that the agents' choice probabilities are based on their subjective value difference $(\Delta \tilde{v})$, whereas their expected random fitness is based on the average expected values of each option ($\pi$) computed under the unbiased regimen for all types of agents (i.e. how much rewards the proportion of the agents choosing one option should actually expect to receive from the physical environment). $\Delta F_{max}$ in Eq.6c serves as a normalisation constant to make sure that $r_{A \to B} \geq 0$; such that the transition rates between the groups will always remain positive (Bladon, Galla et al. 2010), and it is calculated by the following formula:

$$\Delta F_{max} := N \cdot (\pi_L + \pi_R) \qquad (6e)$$

considering the full range of the value space of the dynamically changing, volatile environment in any given generation, which is also the limit of the maximum fitness difference which could be observed between any possible competitors in any generation.

Exact group trajectories were generated by the standard kinetic Monte Carlo algorithm proposed by Gillespie (Gillespie 1976), whereby the population density of group $A$ at time ($t+1$) from time ($t$) is calculated as follows:

$$N_A^{t+1} = (N_A^t - \sum_B N \cdot r_{A \to B}^t + \sum_B N \cdot r_{B \to A}^t) \qquad (7)$$

We simulated between group competitions in different settings (i.e. 4 different $\beta$ values) in 10,000 volatile environments (containing of 100 randomised gambles with different expected value difference ($\Delta v$)). Visual inspection of the Standard Error of Measurement (SEM) margins (i.e. shaded area around the mean trajectories) suggests that the degree of volatility and the behaviour of the strategies across different simulations were mostly comparable. In support for the reliability of the agent-based model we propose here, previous work shows that when competing groups sizes are ≥ 3000 agents, outcome of the kinetic Monte Carlo simulations converge with the trajectories obtained from solving deterministic, mean-field [differential] replicator equations (Traulsen, Claussen et al. 2006) which define the evolution of the system at the infinite population limit.

## 6. Generating intermediate value-based decision-making strategies by simultaneously varying the values of the α (probability weighting) and ρ (risk) coefficients

In a follow-up analysis to the ones shown in Fig.2 B~G, evolutionary fitness of intermediate strategies were tested in an all-out simulation where all strategies competed against each other. These different strategies were generated by varying the values of the α and ρ coefficients. Here, the traditional $\gamma$ coefficient in the 2-parameter probability weighting function was replaced by the α coefficient, which was used to define the original overweighting and underweighting strategies, and the δ coefficient was fixed to its original value (i.e. $\delta$ = 1.05). The parameter space was defined by MATLAB's *linspace* function, whereby 10 possible values of the α and 12 possible values of the ρ coefficient were generated between 0 to 3 and 0.5 to 1.5, respectively. Thus, for every value of the ρ coefficient (i.e. the risk parameter) and at every step of the α (i.e. probability weighting) coefficient, the volatile simulation environment contained 4 competitors (in total 10x12x4 strategies), defined by the following probability weighting equations:

$$\tilde{p} = \exp^{(-\alpha(-\ln(p))^\delta)} \qquad (8a)$$

by which the original overweighting strategy was modified; and

$$\tilde{p} = \exp^{(-(3-\alpha)(-\ln(p))^{\delta})} \tag{8b}$$

by which the original underweighting strategy was modified; and

$$\tilde{p} = 2^{(-(-\log_2(p))^{\alpha})} \tag{8c}$$

by which the original S-shaped strategy was modified; and

$$\tilde{p} = 2^{(-(-\log_2(p))^{(3-\alpha)})} \tag{8d}$$

by which the original inverse S-shaped strategy was modified. The graphical expressions of the output probability weighting functions are given in Fig 4A.

**Supplementary References:**

**Supplementary Figures and Legends:**

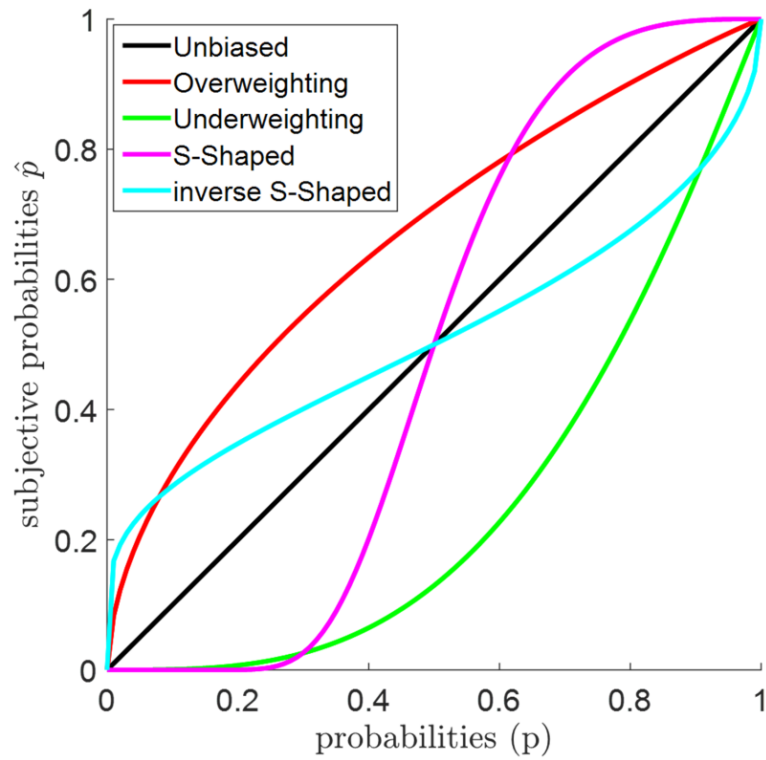

**Supplementary Figure 1. The graphical expression of different probability weighting preferences.** Initial simulations included 5 categorical probability weighting preferences as shown. Probability weighting functions transform raw probabilities on the x-Axis into subjective probabilities on the y-Axis.

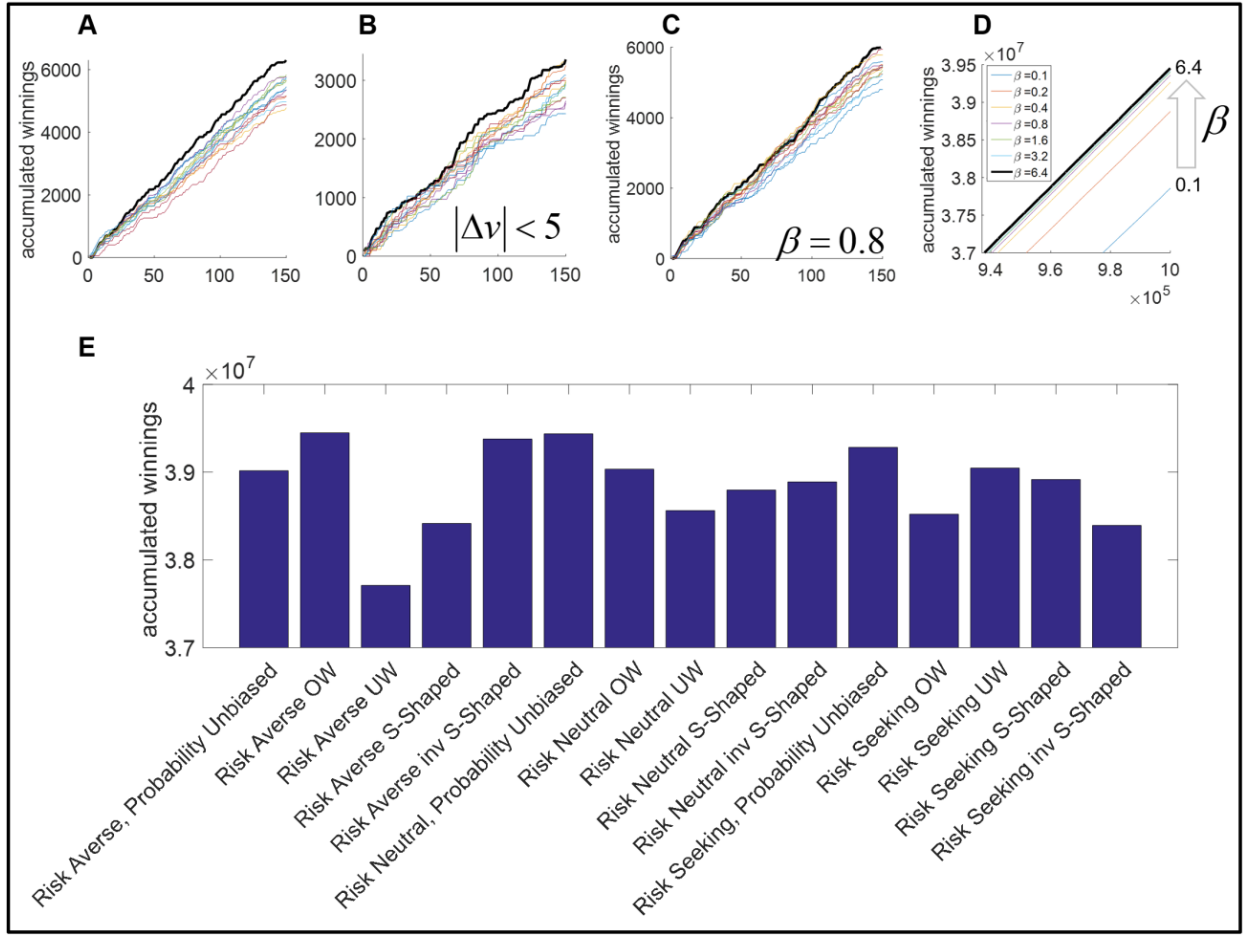

**Supplementary Figure 2. Value-based decision-making under deterministic and stochastic choice models.** The risk neutral strategy with unbiased probability weighting (thick black lines) acquires the highest accumulated rewards in an environment where the expected value difference between the options are **(A)** purely randomised or **(B)** randomised within a limited range. **(C)** The performance of value-based decision-making strategies are negatively affected under the stochastic choice model ($\beta$ = 0.8). Outputs from only the first 150 gambles (x-axes) are shown for demonstration purposes. **(D)** While using the risk neutral strategy with unbiased probability weighting as a template, assigning higher values to the inverse temperature term ($\beta$) in the stochastic choice model shows that accumulated rewards gradually saturate for values of the $\beta \geq 1.6$. **(E)** Accumulated rewards shown for all strategies at the end of 1E6 probabilistic gambles. Accumulated rewards in value-based decision-making is significantly influenced by risk attitude, probability weighting and their interaction (all p<.001).

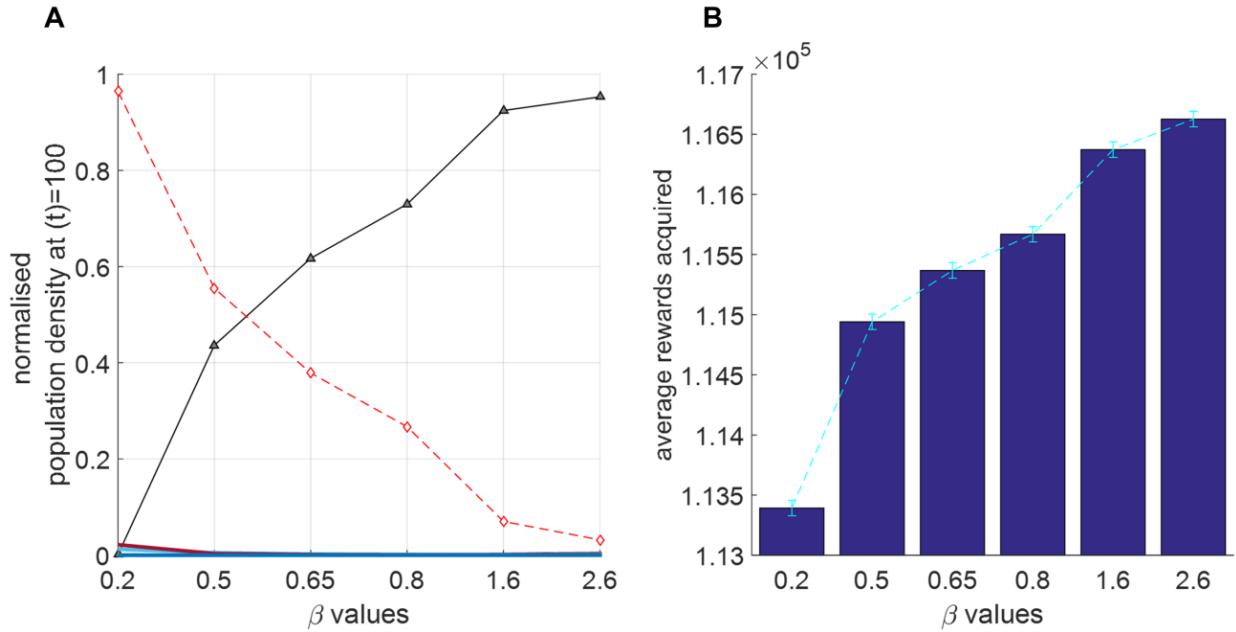

**Supplementary Figure 3. Summary of the evolutionary simulations (A)** showing the changes in the final normalised population density of each strategy at *t* = 100 while increasing values of the β coefficient. Increasing value of the β coefficient reduce the fitness of the risk-seeking strategy (red dashed lines with diamond markers) and augment the fitness of the risk neutral strategy (black line with triangle markers) **(B)** Increasing values of the β coefficient in the stochastic choice model improves the average magnitude of rewards acquired by the population and it gradually saturates. The error bars denote ±1 SEM across 10,000 simulations.

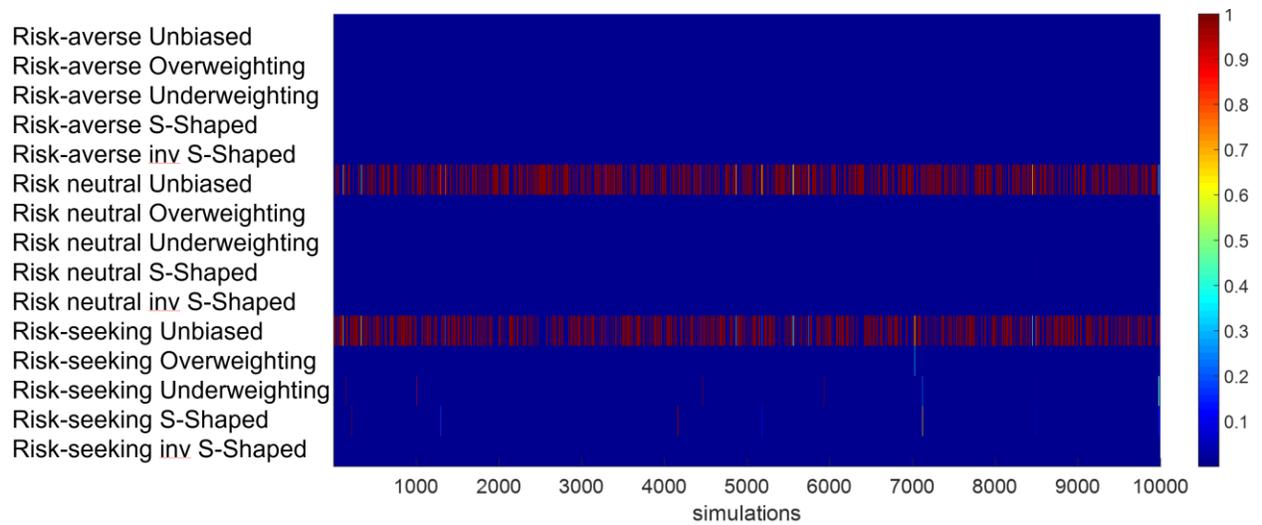

**Supplementary Figure 4. Heatmap summary of the evolutionary simulations when choice stochasticity leads to highest level of competition in the population.** When the inverse temperature term (β) in the sigmoid function is set to 0.55 (see Supplementary Figure 3A), there is strong competition between the risk neutral strategy with unbiased probability weighting and three risk-seeking strategies with different probability weighting preferences. The colour bar represents the normalised population density at the end of each simulation.

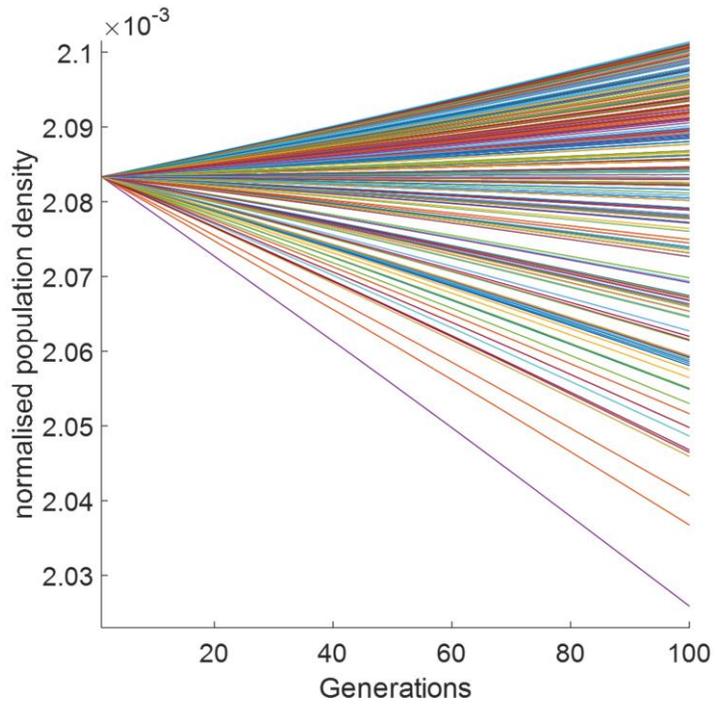

**Supplementary Figure 5. The trajectories of 480 discrete strategies diverge across simulations in 10,000 volatile environments.**

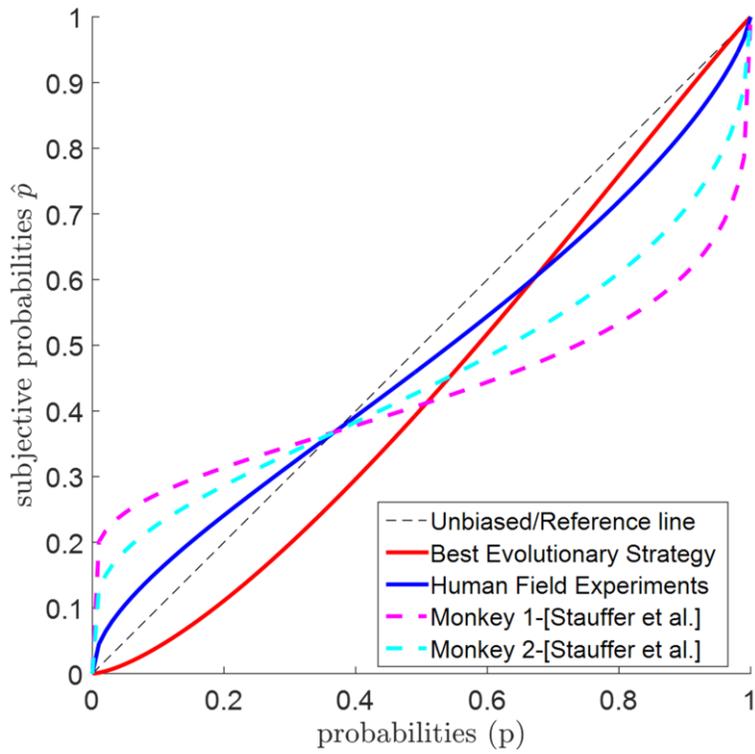

**Supplementary Figure 6. Graphical expression of probability weighting functions which best account for value-based decision-making in humans (Wu and Gonzalez 1998, Tanaka, Camerer et al. 2010) and primates, and how these compare to the trajectory of the probability weighting function of the best evolutionary strategy identified in the current work.**

**link to the Supplementary Video: https://youtu.be/D2_VUXjDRBc**

**Video showing the evolution of normalised population densities for 480 discrete value-based decision-making strategies generated by varying the risk and probability weighting parameters. On a timeline of 3000 generations (i.e. 30 selected randomly from the original 10000 simulation environments) the simulation shows the rise of risk-seeking tendencies. The end of the simulation shows that the model society reaches an equilibrium with only two competing risk-seeking strategies remaining. Note that the analysis provided in Figure 4B discriminates between the best and the second-best strategies across 10000 simulation environments.**